**Authors:**

Pekka Abrahamsson, Karlheinz Kautz

# The Personal Software Process:
# Experiences from Denmark

## Pekka Abrahamsson[1] and Karlheinz Kautz[2]

**[1]Technical Research Centre of Finland, VTT Electronics**

**P.O. Box 1100, FIN-90571 Oulu, Finland**

**pekka.abrahamsson@jyu.fi**

**[2]Copenhagen Business School, Department of Informatics**

**Howitzvej 60, 3., DK-2000, Denmark**

**karl.kautz@cbs.dk**

## Abstract

Software process improvement (SPI) research and practice is transforming from the traditional large-scale assessment based improvement initiatives into smaller sized, tailored initiatives where the emphasis is set on the development personnel and their personal abilities. The personal software process (PSP[SM]) is a method for improving the personal capabilities of a single software engineer. This paper contributes to the body of knowledge within this area by reporting experiences from Denmark. The results indicate an improvement in the effort estimation skills and a significant increase in the resulting product quality in terms of reduced total defect density. The data shows that with relatively small effort (i.e., 10%) used in defect prevention activities (i.e., design and code reviews) almost one third of all defects were removed and consequently the time required for the testing was cut by 50%. Based on this data the use of the PSP method in the software industry is discussed.

## 1. Introduction

In the past 15 years a number of software process improvement (SPI) methods have been introduced. While positive results have been obtained, many of the SPI initiatives fall short of their expectations. In fact, organisations are struggling even in the simplest metrics programs [1]. A realisation that software process is a learning process [e.g., 2] has brought the attention to people-centred process improvement approaches [3]. Thus the emphasis is set on the abilities and competence of the development personnel.

One of the most prominent approaches for the competence development is the personal software process (PSP) method developed by Humphrey [4]. However, only a limited number of the research efforts concerning the PSP are documented. Moreover, software engineering textbooks provide a variety of practical methods to be used in industry. While software professionals seek rational basis for making a decision which method they should adopt, the basis for such a rationalization is completely missing. Methods introduced continue be based more on faith than on an empirical data [5]. There is no quick solution to the problem described. Fenton [5] suggested that only by contributing gradually to the empirical body of knowledge within the specific area of application are we as researchers able to test the basic software engineering hypotheses made. Our principal aim, therefore, is to contribute to the empirical body of knowledge within the area of software engineering and in specific within the area of personal competence development.

The data for this study is obtained from a PSP course held in Copenhagen Business School, Denmark in fall 2001. Research [6, 7] has shown that students are valid representatives for practitioners in industry. We thus believe that this gives valuable insights into the effect of the PSP in general. The results indicate an improvement in the effort estimation skills and a significant increase in the resulting product quality in terms of reduced total defect density. The data shows that with relatively small effort (i.e., 10%) used in defect prevention activities (i.e., design and code reviews) almost one third of all defects were removed and consequently the time required for the testing was cut by 50%.

The paper is organized as follows. The following section provides an overview of the PSP method. This is followed by an introduction to the research setting. The results are presented in section 4 and discussed subsequently in section 5. The paper is concluded with final remarks.

## 2. Overview of the PSP

The PSP was developed by Watts Humphrey [4] to extend the improvement process from an organisation or a project to an individual software engineer. The underlying principle in PSP states that every engineer should do quality work. A high level of quality is achieved through the disciplined utilisation of sound software engineering principles. These principles include a strong focus on the measurement of individual performance. The aim of the PSP is thus to enable software engineers to control and manage their software products as well as to improve their predictability and quality. This is achieved through the gradual introduction of new elements into the baseline personal process in a series of 7-10 small programming tasks. The progression of PSP is shown in Figure 1.

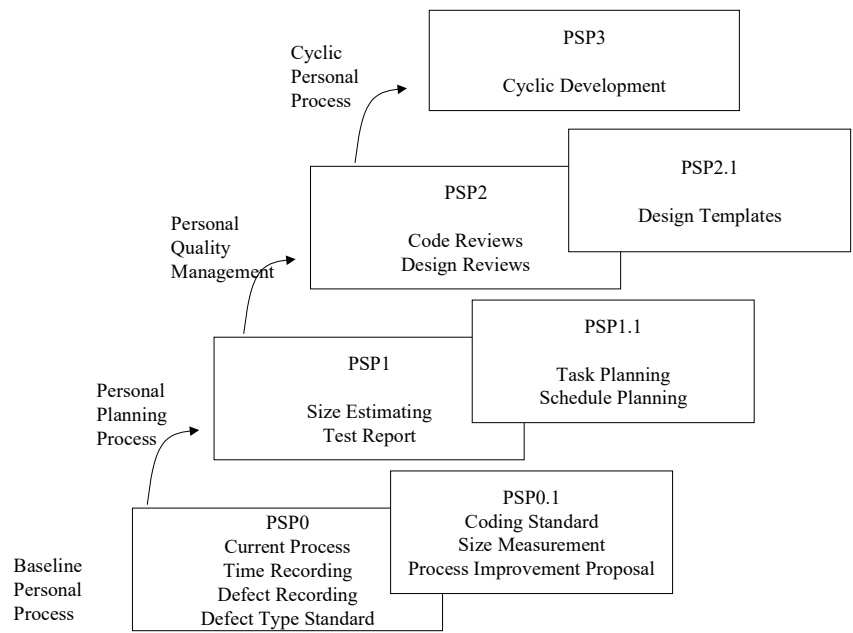

Figure 1. The PSP process development

A student entering a PSP course starts with PSP0, that is, their current process enhanced with time and defect tracking instruments. PSP0.1 extends the personal baseline process to include a systematized coding standard, software size measurement in terms of logical lines of code (LOC) and a personal process improvement proposal mechanism. PSP1 augments the initial process to include the size estimation and a test report practices. PSP1.1 extends the personal planning process to involve a resource planning mechanism. At this level the students become aware of the relationship between program size and use of resources. The size and effort estimations are performed using a Proxy Based Estimation (PROBE) method, where students systematically use the historical data they have collected from the programming exercises.

At PSP2 level the focus is directed towards personal quality management through the introduction of code and design review practices. Students develop their personal defect and design review checklists, based on their historical defect data. PSP2.1 extends the process to include design specifications and analyses. Finally, PSP3 scales up the process from a single module development to larger scale projects. As an outcome, the project is divided in a series of smaller sub projects that are then incrementally implemented.

## 3. Research setting

The PSP data presented in this paper is collected from a PSP course held in Copenhagen Business School in fall 2001. The course was divided in 13 two to three hour lectures, eight programming assignments, two reporting assignments and an exam. Humphrey's [4] book was used as the course book. The PSP 3 level was set as the target for the course. Out of 22 students enrolling to the course, 17 finished the course with a pass grade. Course participants were predominantly fourth and fifth year students. While no specific programming language was enforced, java, C++ and visual basic were dominantly used.

For each assignment, students had a full week to complete the work and submit the results. Disney and Johnson [8, 9] have found that the data collected from a PSP course is often error prone. Thus, in order to ensure the validity of the data collected each assignment was rigorously checked and feedback provided. All data inconsistencies were reported and clarified with the student through email communication. The data collection process was facilitated through the use of electronic documents. Automated data collection tools, however, were not used. Time and defect tracking was thus performed manually in spreadsheet templates.

| Program # | Process | Assignment context: File I/O | Median size (LOC) | Median time (h) |
|-----------|---------|------------------------------|-------------------|------------------|
| 1 | PSP0 | Read/write functions | 90,5 | 4,72 |
| 2 | PSP0.1 | LOC Counter, physical lines | 112 | 4,89 |
| 3 | PSP0.1 | LOC Counter, objects | 87 | 4,52 |
| 4 | PSP1 | Data entry modifications | 151 | 6,01 |
| 5 | PSP1.1 | Basic error handling | 40 | 3,50 |
| 6 | PSP2 | Enhanced error handling, basic calculation functions | 82 | 5,27 |
| 7 | PSP2.1 | Sorting function | 124 | 5,53 |
| 8 | PSP3 | A log file parser for time and defect data | 455 | 14,25 |

Table 1. Programming assignment overview

## 4. Results

The primary goals of the PSP method are three folded. First, it attempts to improve an engineer's ability to estimate the work effort in terms of size and time. Second, the PSP method emphasizes the role of early defect removal by introducing the design and code review techniques. Thirdly, it enables engineers systematically to improve their personal process through the use of process improvement proposals as well as data analysis techniques. The results are explored in terms of these three primary goals. Table 1 shows the details of the programming assignments including the process used, the assignment context or problem area, median[1] size of the assignment in terms of lines of code (LOC) as well as the median time used for the development of module size programs. The data presented in the following subsections is systematically grouped according to the major PSP levels, Table 2.

---

[1] Median value shows the midpoint in a data set. This means that 50% of data points are below and 50% are above the median value. Median is more useful for small data sets than the average value when the data points are not equally distributed.

| PSP Level | Programs # | Number of cases |
|-----------|-----------|-----------------|
| PSP0 | 1, 2, 3 | 52 |
| PSP1 | 4, 5 | 34 |
| PSP2/PSP3 | 6, 7, 8 | 47 |

Table 2. The data used in the study

By pooling the data in logically coherent sets - such as the major PSP levels - the analytical validity of the analysis is increased. Thus, the first three programming assignments belong to the PSP0-level, the next two assignments belong to the PSP1-level, and finally the last three assignments belong to the PSP2/PSP3 –level. Table 2 also shows the number of cases, i.e. assignments, belonging to each of the PSP levels.

4.1. Size and effort estimates

In the PSP method, size estimation provides the basis for an effort estimate. The size measure that is used is lines of code. PSP research has repeatedly demonstrated that LOC correlate reasonably well with the development effort. Estimates are based on students' personal data collected from the previous assignments. At PSP0 level the size estimate may thus vary a great deal but this variation should stabilize within a 25% error margin at PSP2/PSP3 level [10]. PSP research argues that a similar trend should also be found concerning the effort estimates even though individual differences may exist. A box plot[2] diagram of the development of the size and the effort estimation accuracy is shown in Figure 2.

---

[2] A box plot diagram visualises the 5 number summary of a data set. Median value is the line in the shaded box area. Q1 (first or lower quartile) shows the median of the lower 50% of data points. Q3 (third or upper quartile) shows the median of upper 50% of data points. The minimum value indicates the lowest and the maximum the highest values in the respective data sets.

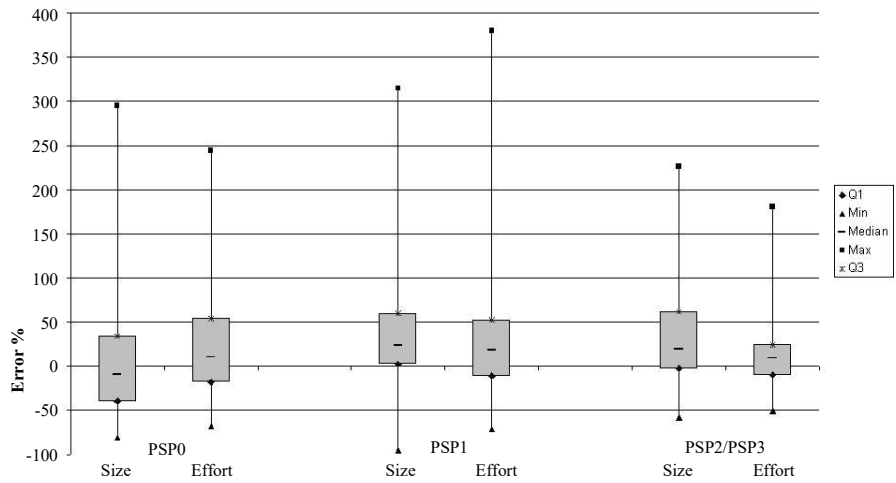

Figure 2. Size and effort estimation accuracy

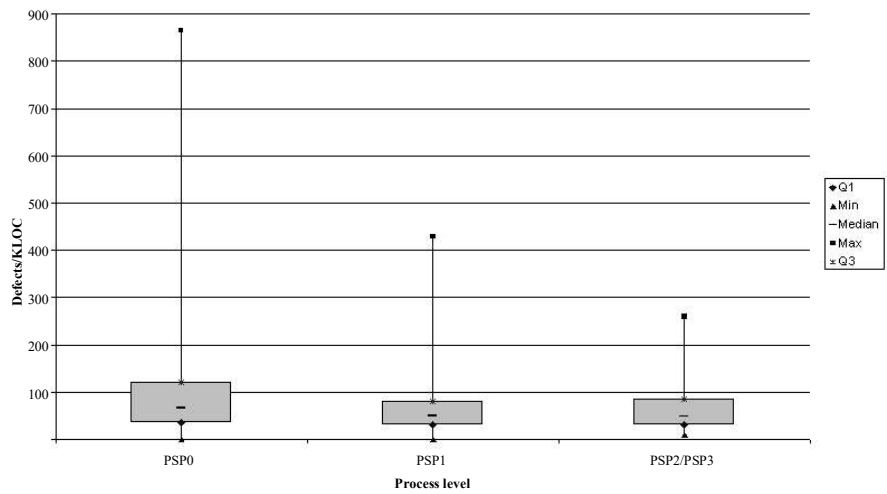

Figure 3. Overall defect density

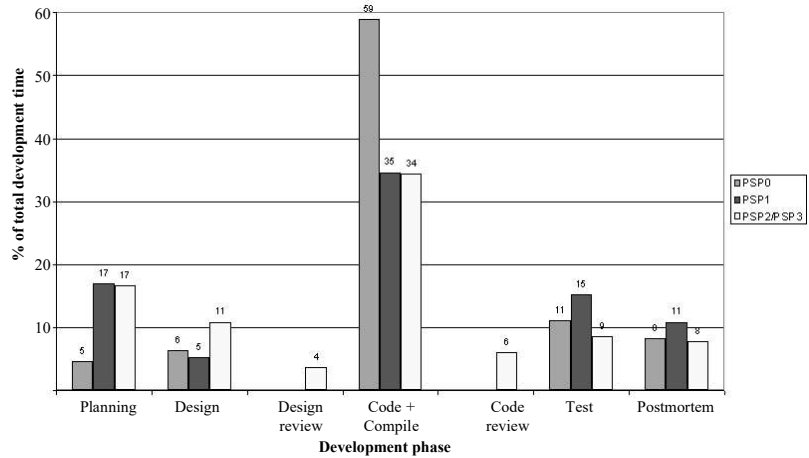

Figure 4. Change in effort distribution

While the data shows no significant improvement in size estimation abilities, the effort estimation error range stabilized within the 25% error margin indicating a significant improvement when compared with the PSP0 and the PSP1 levels.

## 4.2. Product quality

The PSP method emphasizes the role of early defect removal as a cost-effective way to increase the quality of the resulting product. Design and code reviews are the main techniques introduced. Hayes and Over found [11] that the overall defect density was reduced by a factor of 1.5. Figure 3 shows the development of overall defect density over the three main PSP phases.

The data shows that the median was reduced from 67 (PSP0 level) to 48 (PSP2/PSP3 level) defects/KLOC. This indicates an improvement by a factor of 1.4. Similarly, defects found in the test phase were reduced from 10 to 5 defects/KLOC indicating an improvement by a factor of 2.1. Table 3 shows the total number of defects removed in each development phase.

|  | Pla | Design | DR | Code | CR | Tes |
|---|---|---|---|---|---|---|

|  | n |  |  | + compile |  | t |
|---|---|---|---|---|---|---|
| PSP0 | 0 0% | 5 1% | - | 346 69% | - | 152 30% |
| PSP1 | 0 0% | 0 0% | - | 154 70% | - | 65 30% |
| PSP2/3 | 0 0% | 8 1% | 33 6% | 325 57% | 122 21% | 83 15% |

Table 3. The number and % of defects removed

The defect data shows no change in defect removal profile between the first two PSP levels. The use of design and code reviews decrease the % of defects removed in both the implementation (i.e., code and compile) and in the test phase.

## 4.3 Effort distribution

The PSP method guides the development of module level programs through a series of process scripts for each of the development phase. These process scripts define the entry and the exit criteria, process activities and the outcome of each phase. Each PSP level introduced incorporates new elements into the process such as an explicit method for the size and the effort estimation (PSP1) and the code and the design reviews (PSP2). Thus, when the method is being learned the effort distribution should change by lessening the time used for the implementation phase (code and compile) and by increasing the time used for other phases. While the code and the design review mechanisms place emphasis on the early defect removal, the time used for testing should decrease. Figure 4 shows the median development of the effort distribution over the PSP levels.

As expected, the most significant change in the effort distribution is the amount of time spent in the implementation phase. However, there is no change between the PSP1 and PSP2/PSP3 levels in this regard. At the PSP2/PSP3 level the reviews are introduced. Based on data on 47 module level programs the effort used for the reviews altogether is 10%. The defect data shows that with this 10% effort 27% of all defects injected were caught. As a result of this, 50% less defects were found in the test phase, which may explain the reduction in time used for testing. The effort spent in the data summary and the analysis, i.e. postmortem, phase shows no significant change.

## 4.4. Student feedback

Feedback was collected from each assignment relating to the problem context and to the current PSP process. When establishing the baseline process (i.e., PSP0) the students found it beneficial to understand their effort distribution over different software development phases. Detailed time tracking also made the students realize how fragmented their work is, i.e. there are lot of interruptions that distract them from the development work.

"[It was positive] that you actually get a picture of much time you spend on interruptions and on the different [development] phases."

As suggested by Humphrey [4] the PSP method is better understood when the course participants are well versed in the programming language they use to implement the assignments. This enables the students to concentrate on the process experimentation. Some of the course participants faced difficulties due to a lack of adequate programming skills:

"It is not the PSP procedures but my programming skills that worry me."

"I finally got the hang of pointers in C++. Now I can't understand why I thought that it was difficult to begin with."

The PSP method is claimed to enable a software engineer to gain control over the process and then to improve it in a systematic way. It was found that the PSP enables the students to identify the targets for improvement rather efficiently, and, more importantly, the students are able to provide different proposals on how to solve the problem.

"If I remove the worst cases from the estimation data, I actually get an excellent correlation, but using this doesn't make the estimation any better. [...] The problem therefore must lie in the filling out the [Probe Method Template] or the way I regulate the PMT result."

The PSP method is an experimentation oriented approach to software development. In the learning phase, the students use a wide range of different techniques and approaches for specific tasks. This enables them to judge the value of each method for the future use. As an example, a student found that while the code review can be done efficiently, the pair-review technique is better for the design phase.

"When doing the design review alone it is VERY difficult to find errors. This time my friend joined the review and came up with a much better and more object-oriented design."

When automated data collection and analysis devices are not used, the process becomes rather heavy in terms of number of different documents that need to be managed. While this is acceptable in academic setting, industrial PSP users need efficient tools to support their practice.

"It is very time consuming and very frustrating to look at all [the] documents during the process."

Students also expressed dissatisfaction with the fact that the PSP course requires much more effort than software engineering courses in general.

"I am glad that these assignments have finally come to an end… the workload has been tremendous, not at all in relation with the small ratio this course [counts in] the final exam papers."

While the PSP method is learned in a very practical manner, it makes a significant contribution to students' general knowledge and understanding of software engineering at a personal level.

"It has been nice to experience how a software process […] can be carried out. It is much different from my earlier experiences. […] This one has an advantage [in comparison to] others, since it makes […] the process visible to its user. Afterwards […] it is possible to evaluate the process on the basis of facts and not feelings."

5. Discussion

The basic promises that the PSP method proponents claim – i.e., increased process visibility, better control over the work and the systematic improvement framework – are supported by the results collected in this study.

The results indicate that while the size estimation ability did not show significant improvement, the ability to estimate the required work effort did improve. Improvement was also identifiable in terms of overall and test defect density. It should be noted that in a classroom setting little improvement in these skills is generally expected. Ability to estimate is dependent on the quality of the historical data collected, which in the course setting is questionable to some extent. Moreover, even when the PSP estimating techniques are used over an extensive period of time – e.g., five years – some fluctuation in the size and effort estimation accuracy still may exist. [12]. However, the ability to improve already in the learning phase works as a motivational factor in regard to the post-course use of the PSP method. We support here Prechelt and Unger's [13] argument who maintain that the potential benefits of the method are often not directly observable during the course and they do not necessarily realize automatically even after the course. This may be due to the fact the most software produced in industry is what can be called domain dependent software [14].

The software produced in the PSP course is domain independent and when applied in an industrial setting, the method needs to be adjusted to fit the environment. This adjustment, again, takes time and effort and lessens the visibility of observable improvements.

While our results are not new or surprising, they add to the much-needed body of knowledge within the area of software engineering and especially within the area of software engineers' competence development. Wohlin [15] suggested that the PSP course offers a suitable environment and context for conducting experimental studies to test many of the software engineering hypotheses made. Our findings support his claim in this regard. However, this requires a rigorous and, to some extent, automated data collection process where the validity of the data can be efficiently verified.

Based on the data obtained in this and other similar studies, we – as researchers – should be able to answer whether the software industry should invest in the PSP method or should other means rather be explored in hope for better benefits. Research has shown that many of the large scale software process improvement (SPI) initiatives often fall short of their intended goals [e.g., 16] and the role of SPI department is often reduced to basic support activities with little strategic importance [17]. The basic problems of software engineering, however, have not been solved. Emerging methods such as xP [18] place emphasis and reliance on the abilities of a single software engineer but is not clear on how to develop and maintain such a competence. The PSP method is essentially about individual software engineer's ability to learn to control and to develop his/her own processes. Only after having explored different techniques an engineer is able to decide upon the most effective solution. Moreover, the use the PSP indicates increased personal responsibility for quality and productivity improvements [19]. While the software engineering research is keen in introducing new and enhanced methods, often the evaluation of existing ones is limited [20]. Best results in industry have been obtained when the PSP method is tailored to the operating context by taking into account the culture and the project management practices [e.g., 21, 22]. Thus, we claim that only by enabling the software engineers to develop and maintain their professional competence, significant improvements in quality and productivity are to be reached. The PSP method contains all necessary elements in such a development process. We argue that the role of universities and other institutions is important in this regard. As a part of solution to bring much-needed rigor in the software development, universities should consider including the elements of the PSP method or the method itself into the course curriculum.

6. Conclusion

This paper reported PSP experiences from Denmark. The data for this study was obtained from a PSP course held in Copenhagen Business School, Denmark in fall 2001. The results did indicate an improvement in the effort estimation skills and in the resulting product quality in terms of reduced total defect density. The effectiveness of the defect prevention activities (i.e., design and code reviews) were demonstrated at a personal level based on the data. Finally, it was suggested that other universities and institutions should consider incorporating the elements of the PSP method into the course curriculum due to its focus at the personal level, which is the source for the most long-standing improvements.

Acknowledgements

The authors would like to thank the students participating in the PSP course for their effort in collecting the PSP data and for their valuable comments throughout the course, and the four anonymous reviewers for their comments on the early version of the paper.